\documentclass[a4paper,11pt]{article}
\pdfoutput=1 

\usepackage{jinstpub} 

\title{\boldmath Energy scale calibration of KEDR detector tagging system}

\author{V.V. Kaminskiy,}
\author{N.Yu. Muchnoi and}
\author{V.N. Zhilich}

\affiliation{Budker Institute of Nuclear Physics SB RAS,\\ Novosibirsk, Russia}
\affiliation{Novosibirsk State University, \\Novosibirsk, Russia}

\emailAdd{v.v.kaminskiy@inp.nsk.su}

\abstract{The tagging system of the KEDR detector is a symmetrical focusing magnetic spectrometer for electrons and positrons scattered at small angles; it is embedded into the lattice of the VEPP-4M collider and intended for two-photon processes study. It measures scattered electron energy with resolution $\Delta E/E_0=0.03\%...0.6\%$ ($E_0$ is the beam energy). For precise energy scale calibration two methods are used: tagging of bremsstrahlung electron/positron by the photon energy measured by BGO calorimeter, and direct calibration using Compton backscattering spectrum edge. This report covers design and current status of the calibration system.}


\arxivnumber{1705.10114} 

\proceeding{Instrumentation for Colliding Beam Physics\\
  27 February - 3 March, 2017\\
  Budker Institute of Nuclear Physics, Novosibirsk, Russia}

\begin{document}
\maketitle
\flushbottom

\section{KEDR detector tagging system}
\label{sec:ts}

Two-photon processes $e^-e^+ \rightarrow \gamma^*\gamma^* e^-e^+ \rightarrow X e^-e^+ $ study is an essential part of the KEDR detector and the VEPP-4M collider (at BINP SB RAS) physical program \cite{KEDR}. For this purpose the KEDR detector has a unique tagging system (TS) for electrons and positrons scattered at small angles after interaction \cite{TS}. The invariant two-photon mass can be calculated:
\begin{equation}
W_{\gamma^*\gamma^*}^2\approx 4 \omega_1 \omega_2 \approx 4(E_0-E_-)(E_0-E_+)\,,
\label{eq:2g_mass}
\end{equation}
where $\omega_1$, $\omega_2$ are virtual photons energies, $E_0$ is the beam energy, $E_-$, $E_+$ are scattered electron and positron energies.

TS is a symmetrical focusing magnetic spectrometer embedded in the VEPP-4M lattice. It consists of 4 bending magnets, 4 quadrupole lenses, 3 solenoids and also sextupoles, beam orbit correctors and other elements. Scattered electrons (or positrons, briefly, SE) initially go along the equilibrium beam; then they are deflected by dipole magnets and registered by coordinate detectors (drift tube hodoscope and GEM detector). Two quadrupole lenses focus SE at some distance depending of its energy; dipole magnets deflect the focus point from the equilibrium beam. For optimum resolution 8 coordinate detectors are placed at the focus curve. The spectrometer energy resolution varies from 0.03\% to 0.6\% of the beam energy, and two-photon system invariant mass resolution varies from 3~GeV to 20~GeV in the beam energy range of 1.5--5.0~GeV.

\begin{figure}[htbp]
\centering
\includegraphics[width=\textwidth]{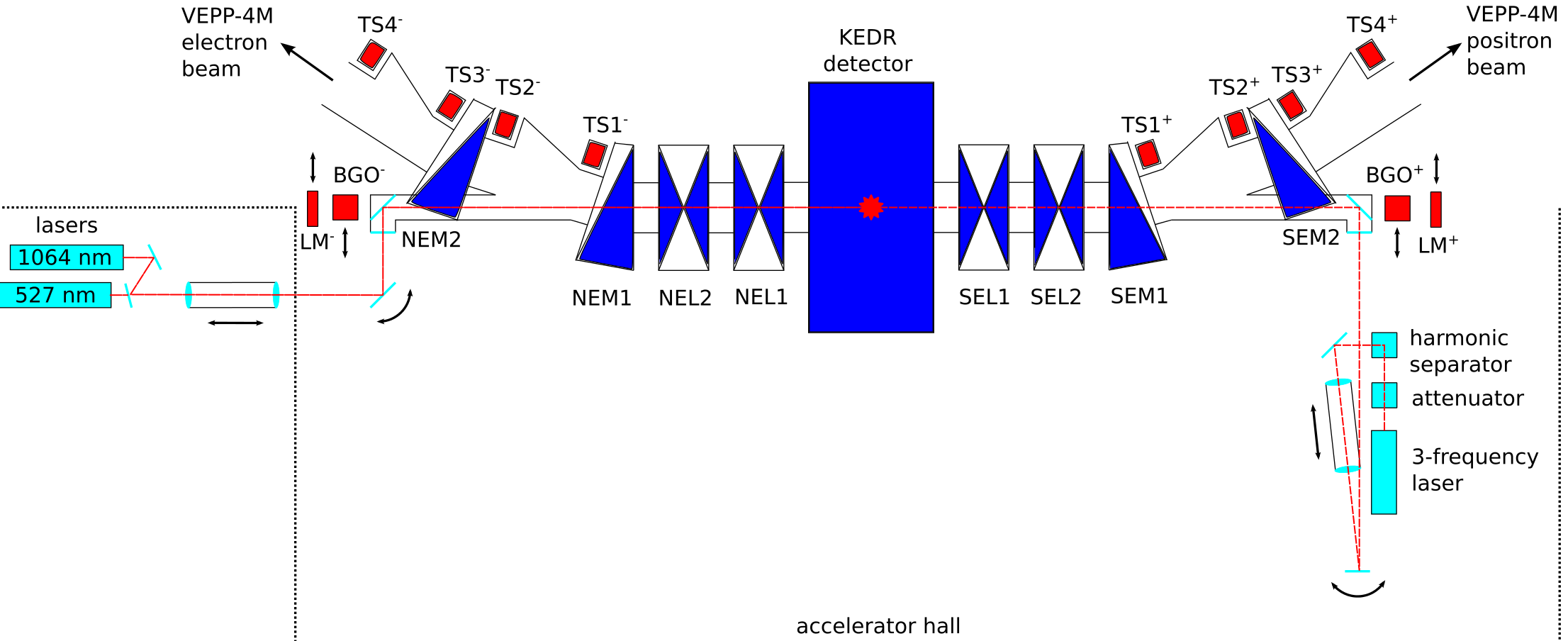}
\caption{Layout of the KEDR detector tagging system. NEM, SEM are bending magnets; NEL, SEL are quadrupole lenses; TS are coordinate detectors; BGO are calorimeters; LM are luminosity monitors.}
\end{figure}
In common case, the horizontal coordinate $X$ depends on the SE energy $E$:
\begin{equation}
X = A \frac{E_0}{E} + B\,,
\label{eq:scale}
\end{equation}
where $E_0$ is the beam energy, which is known at the VEPP-4M collider with accuracy better than $10^{-4}$ (below 2~GeV); $A$ and $B$ depend on bending magnets field, the TS modules geometry, the beam orbit and fields of the other magnets. So, if one knows $A$, $B$, $E_0$ and measures $X$, one can calculate the SE energy. For a high precision, the TS energy scale should be calibrated using a physical processes.

\begin{figure}[htbp]
\centering
\includegraphics[width=0.5\textwidth]{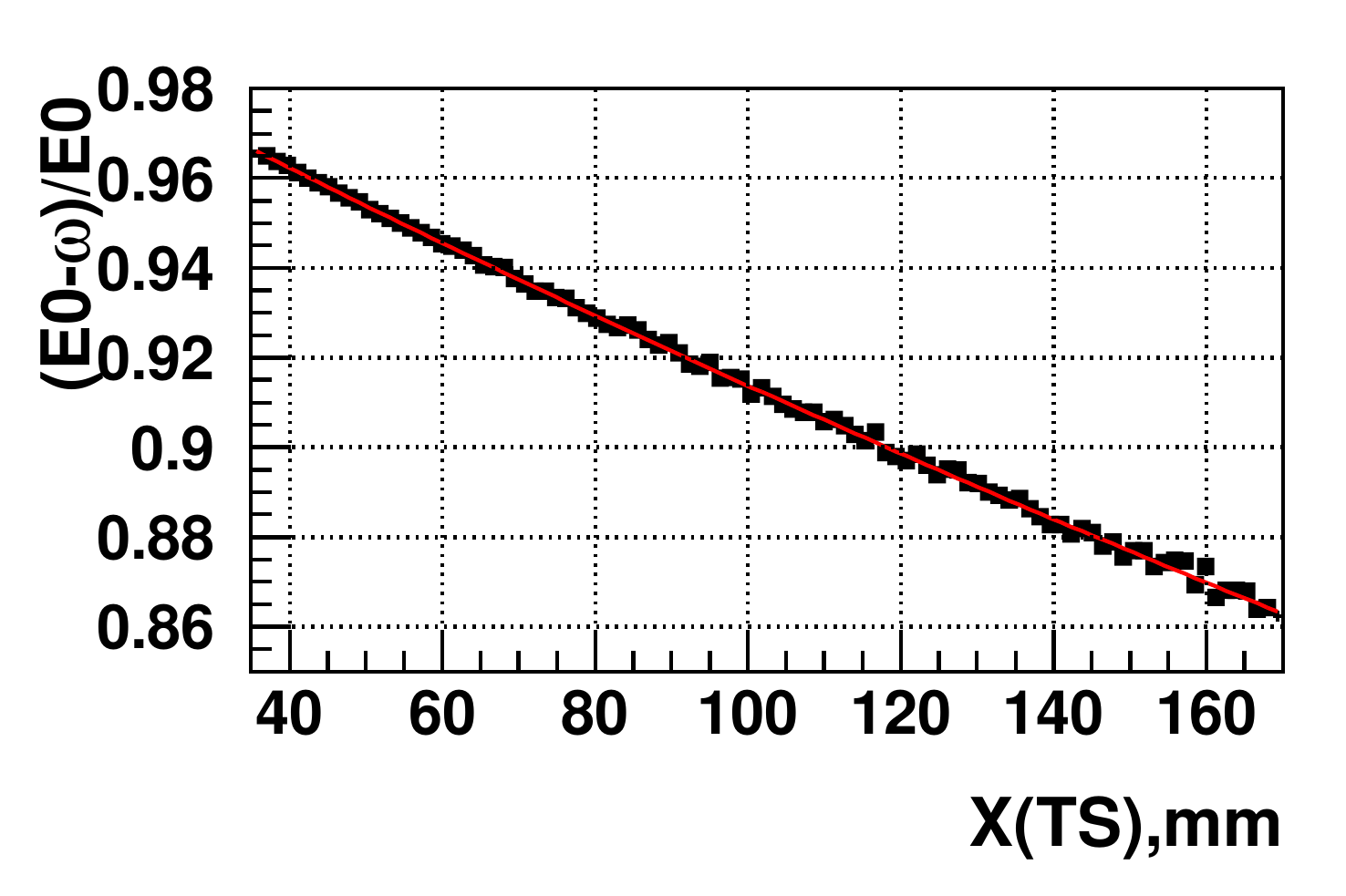}%
\includegraphics[width=0.5\textwidth]{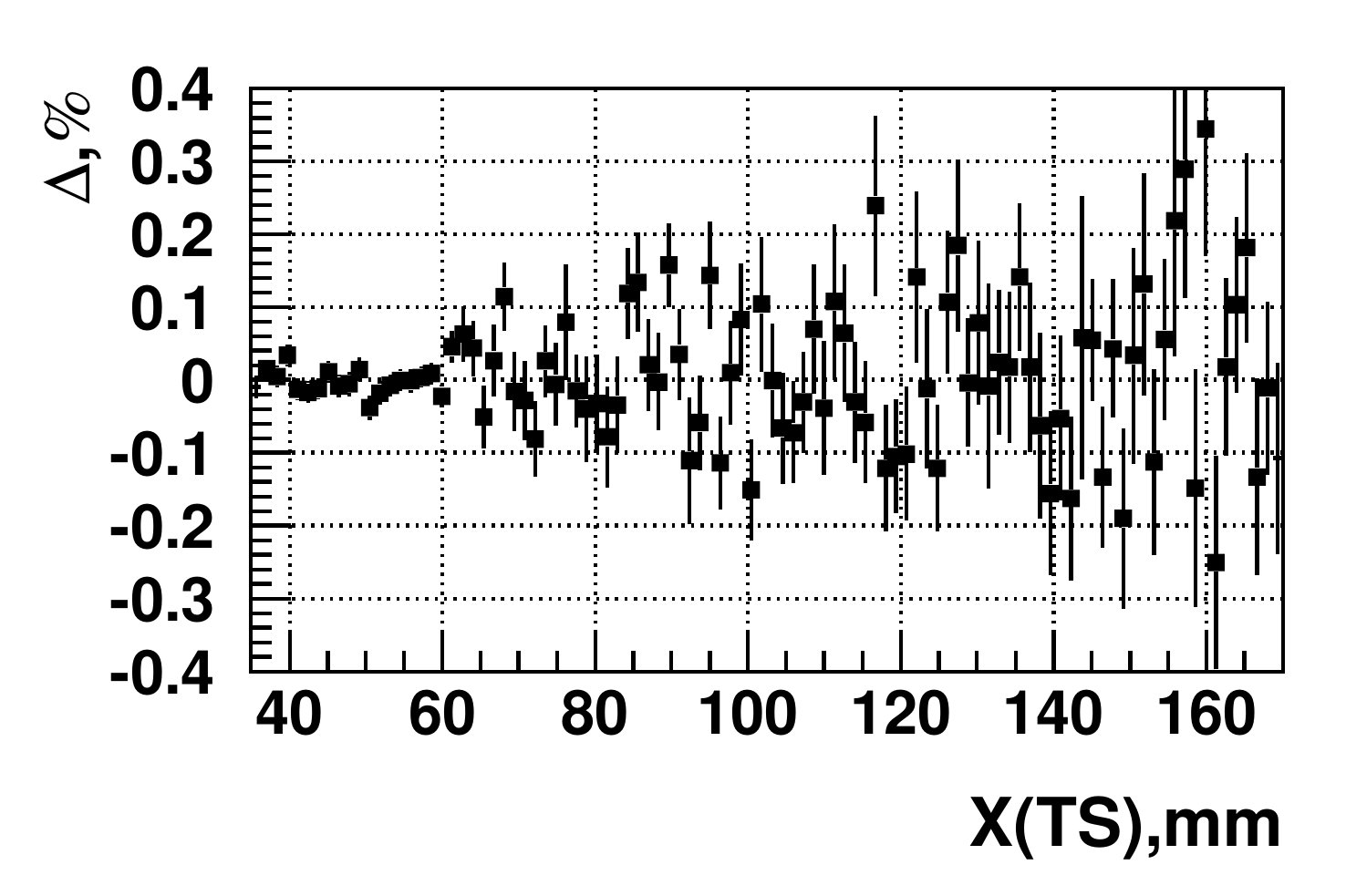}
\caption{At right: a calibration of the TS4$^-$ coordinate module using the BGO calorimeter. The graph is fitted with a function $(E_0-\omega)/E_0 = A/(X - B)$ (it is derived from equation~\ref{eq:scale}). At left: a difference between $(E_0-\omega)/E_0$ and the fitting curve which can be considered as the accuracy/nonlinearity of the calibration (larger points spread at higher coordinate is due to lower event count).}
\label{fig:tagging}
\end{figure}

For the TS calibration two methods are used. Single bremsstrahlung (SB) electron energy is a difference between the initial beam energy and a photon energy $\omega$: $E = E_0 - \omega$, so, SE energy tagging by a photon energy is applied \cite{Zhilich_TS_calibr} (see Figure~\ref{fig:tagging}). The TS has two BGO calorimeters for a photon energy measurement. The calorimeters are calibrated using narrow spectrum edges from SB and Compton backscattering (see Figure~\ref{fig:edges}). Other approach uses coordinate spectrum edge of Compton backscattered electrons/positrons.

\section{Compton backscattering}
\label{sec:cbs}

\begin{figure}[htbp]
\centering
\includegraphics[width=0.9\textwidth]{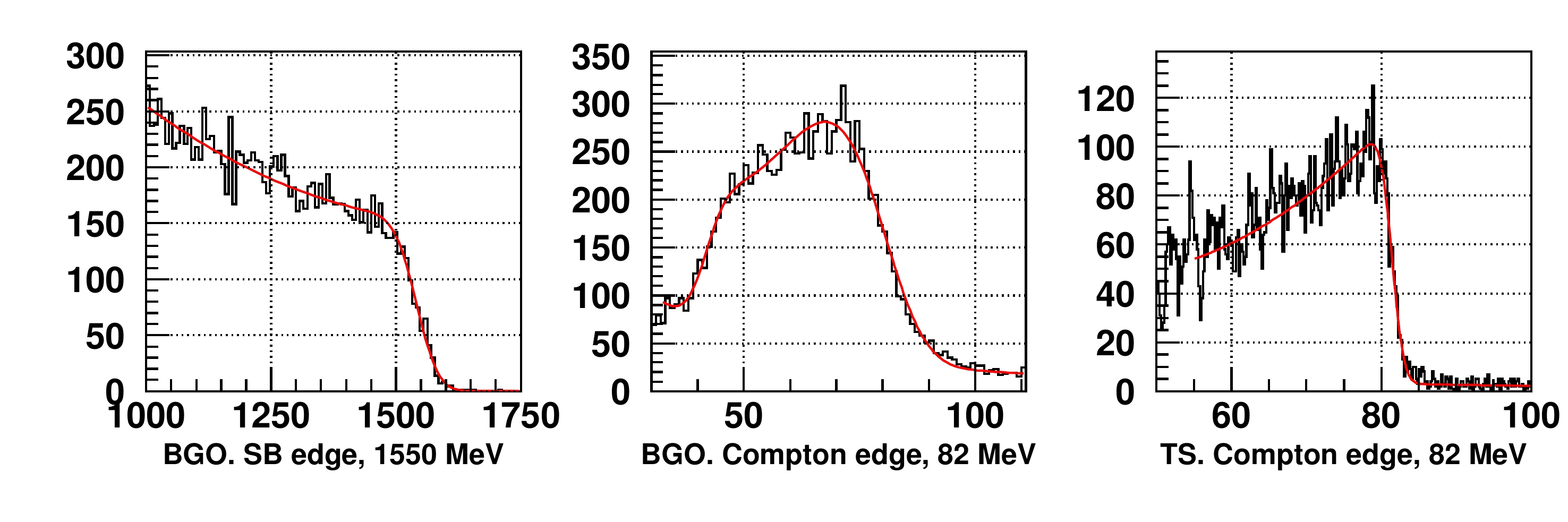}
\caption{Experimental spectrum edges for Compton backscattering and single bremsstrahlung. $E_0=1500$~MeV, \mbox{$\omega_0=2.33$~eV}. First plot from left: SB photon spectrum edge in the BGO calorimeter. Second: CBS photon spectrum edge. Third: electron CBS spectrum edge, X axis is energy {\sl lost} by electron: $E_0-E$, which is converted from coordinate.}
\label{fig:edges}
\end{figure}

Compton backscattering (CBS, or inverse Compton effect) is an inelastic head-on interaction of a low-energy photon ($\omega_0=1.2...4.8$~eV) and an ultra-relativistic electron (or positron) ($E_0=1.5...5.0$~GeV). When the photon scatters at zero angle (in the direction of the initial electron and backwards to the initial photon) the scattered photon energy is maximal and the scattered electron energy is minimal:
\begin{gather}
\omega_{\text{max}} = \frac {E_0 \lambda} {1 + \lambda} \approx 4 \gamma^2 \omega_0\,, \qquad
E_{\text{min}} = E_0-\omega_{\text{max}} = \frac{E_0} {1+\lambda}\,, \\
\lambda = \frac{4 \omega_0 E_0 } {m^2}\,,\nonumber
\end{gather}
where $m$ is an electron rest energy ($c=1$), $\gamma=E_0/m$. So, the energy spectrum of scattered photons is almost flat and extends from zero to a narrow edge at $\omega_{\text{max}}$. The scattered electrons spectrum looks like photons spectrum reflected relatively to the beam energy. See energy and coordinate spectra at the Figure~\ref{fig:edges}.

\begin{figure}[htbp]
\centering
\includegraphics[width=0.7\textwidth]{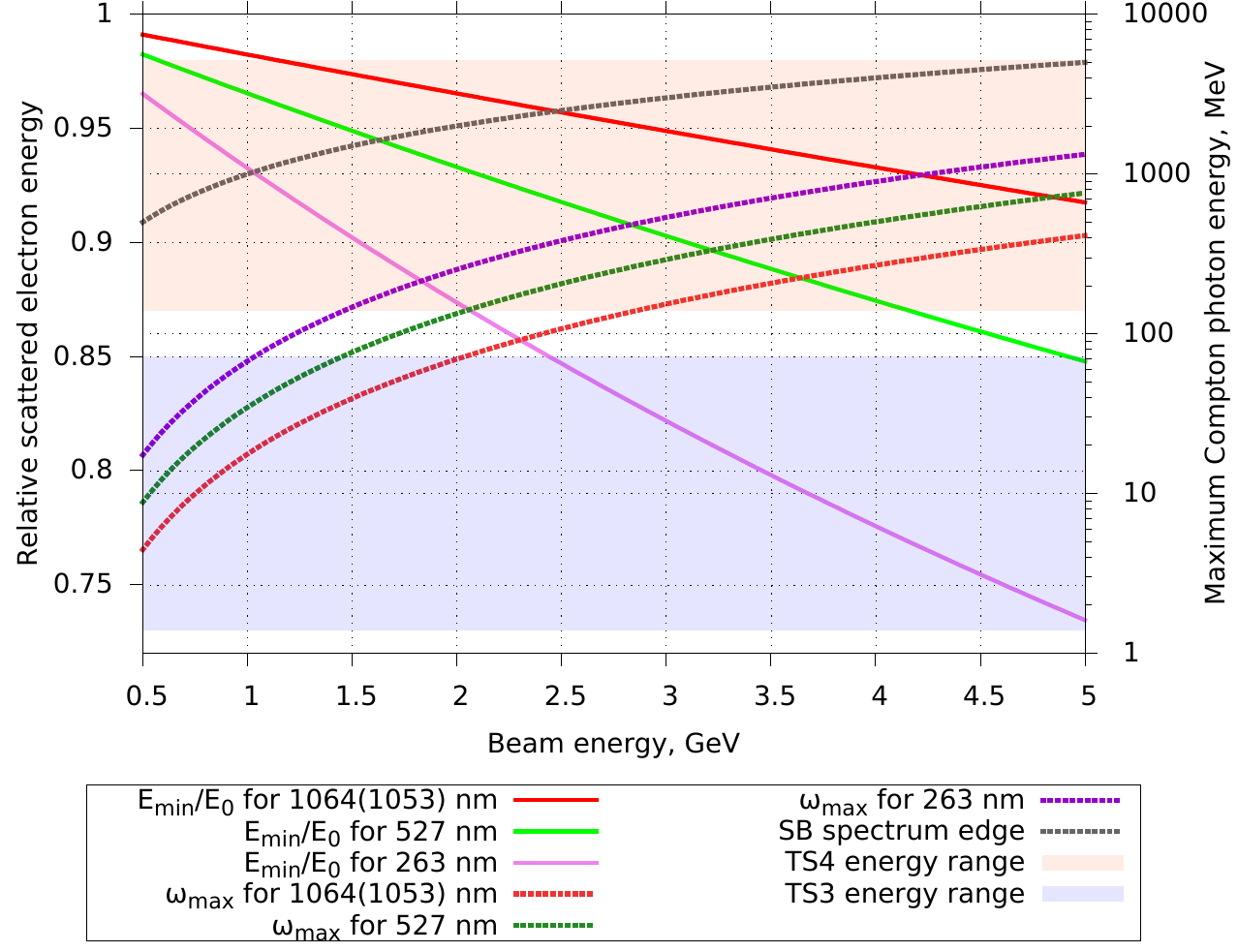}
\caption{Possibilities for the TS calibration: coordinate detectors and BGO calorimeters.}
\label{fig:calibration_possibility}
\end{figure}

So, if one provides the interaction of the monochromatic laser radiation with the electron beam and measures the transverse coordinate of the scattered
electrons with minimum energy (Compton edge), one can match the coordinate and the energy. For a proper definition of $A$ and $B$ coefficients (see expression~\ref{eq:scale}), one needs two spectrum edges, two laser wavelengths. With the lasers used in this work, TS4$^{\pm}$ modules can be calibrated in the beam energy range of 1.5--4.0~GeV (see Figure~\ref{fig:calibration_possibility}). It should be mentioned that most SE created in $\gamma\gamma$ events are registerd in TS4$^{\pm}$ modules due to a low-energy dominated shape ($1/\omega$) of the virtual photons spectrum. Also the spectrum edge of Compton photons is used for the calibration of BGO calorimeters in the energy range of 40--1000~MeV within the same beam energy range (see Figure~\ref{fig:calibration_possibility}).

\section{Optical system}
\label{sec:optics}

Solid-state pulsed lasers with a pulse duration less than 10~ns are used, so, Compton interaction occurs within 2~m around the center of the KEDR detector. A pulse repetition rate is 1000~Hz; it is generated from the signal of the VEPP-4M beam revolution frequency (819~kHz). Two lasers work at the electron side of TS: 1064~nm (Nd:YAG) and 527~nm (Nd:YLF with frequency doubling). Three-frequency Nd:YLF laser works at the positron side of TS: 1053~nm, 527~nm and 263~nm (4th harmonic currently is not used for TS calibration).

At the electron side two laser beams are mixed into one beam using ordinary and dichroic mirrors. The beam is focused by a motorized expander (two lenses in a tube with a variable gap between them) at some distance (approximately 22~m). Then the laser beam goes to a motorised two-axis mirror, which allows to adjust a transverse position of the laser beam in the interaction region. The laser beam is inserted into the vacuum chamber through a fused silica window and is reflected by a static mirror towards the electron beam.

At the positron side the laser beam is attenuated by a motorized polarizing prism. Then three wavelengths of the laser are separated by a motorized dispersing prism. Being reflected by a static mirror, the chosen beam is focused by a motorized expander, positioned by a movable mirror and inserted into the vacuum chamber similarly to the electron side.

A special Compton rate monitor is implemented to adjust the interaction. A special software controls parts of optical system: step motor controllers, delay lines, commutators, etc., and also it controls the whole installation. The software provides long-term (weeks) fully automated experiment. See more about optical system and its control in \cite{TS_calibr}.

\section{Energy scale calculation}

Between physical calibrations, the TS energy scale is calculated using the model of VEPP-4M. 
This model generates particles of SB and CBS and simulate their propagation through the \mbox{VEPP-4M} lattice. The model uses measured currents, fields and the beam orbit within experimental runs (60...120~minutes). Coefficients $A$ and $B$ from equation~\ref{eq:scale} are calculated for 8 modules.

This model calculate relative changes of the energy scale. For calculated coefficients $A_{\text{c}}$ and $B_{\text{c}}$ a correction is made: $A\rightarrow A_{\text{c}} C_A$, $B\rightarrow B_{\text{c}} + \Delta B$. $C_A$ and $\Delta B$ are found from runs with experimental calibration: $C_A = A_{\text{e}}/A_\text{c}$ and $\Delta B = B_{\text{e}} - B_\text{c}$. When comparing calculated and measured Compton edge positions for long term, one can see that the model repeats position jumps and drifts after the experiment within 0.03\% of the beam energy. We plan to improve the accuracy to the theoretical limit of 0.01\%--0.02\%.

\section{Conclusions}

Facility for the energy scale calibration of the KEDR detector tagging system was constructed: laser and optical system with two wavelengths for electron and positron beams, and control system. The facility works continuously and automatically. A method for scattered electron trajectories in the TS was created. It is based on measurement of parameters of the VEPP-4M collider magnetic elements and allows to calculate the TS energy scale with a reasonable accuracy. Work on improving accuracy and stability is going on.

\acknowledgments

This work has been partially supported by Russian Foundation for Basic Research (project No. 15-02-09016 A).

\bibliographystyle{JHEP}
\bibliography{ts_calibration}

\end{document}